\documentclass{aa}  

\def\ergcms{erg~cm$^{-2}$~s$^{-1}$ }
\def\nh{N$_\mathrm{H}$}
\def\cm2{cm$^{-2}$}
\def\ra{RA$_{{\mathrm{J}}2000}$}
\def\dec{DEC$_{{\mathrm{J}}2000}$}
\def\integral{{\it{INTEGRAL}}}
\def\swift{{\it{Swift}}}

\usepackage{color}
\definecolor{red}{rgb}{0.7,0,0}
\definecolor{blue}{rgb}{0,0,0.7}
\def\correc#1{{#1}}

\usepackage{graphicx,epsfig,epsf}
\usepackage{longtable}
\usepackage{natbib}
 \usepackage[figuresright]{rotating}
\usepackage{txfonts}
\begin{document}
   \title{\swift\ follow-up observations of INTEGRAL sources of unknown nature}

   \author{J. Rodriguez
          \inst{1}
          \and
          J.A. Tomsick \inst{2}
	  \and
          S. Chaty\inst{1}}

   \offprints{J. Rodriguez}

   \institute{Laboratoire AIM, CEA/DSM - CNRS - Universit\'e Paris Diderot, DAPNIA/SAp,
 F-91191 Gif-sur-Yvette, France\\
              \email{jrodriguez@cea.fr}
         \and
             Space Sciences Laboratory, 7 Gauss Way,
University of California, Berkeley, CA 94720-7450, USA\\
             }

   \date{}

% \abstract{}{}{}{}{} 
% 5 {} token are mandatory
 
  \abstract
  % context heading (optional)
  % {} leave it empty if necessary  
   {Since its launch in 2002, \integral\ has discovered many new hard 
X-ray sources. A lot of them still lack sufficient positional accuracy, 
for finding counterparts at other wavelengths. Their true 
nature is, therefore, still unknown.}
  % aims heading (mandatory)
   {The goal of this study is to give an accurate X-ray position for 
12 of these sources so as to further identify their counterpart
at optical, infrared, and radio wavelengths, and to unveil their true nature. We 
also make use of the X-ray spectral parameters to tentatively distinguish 
between the various possible types.}
  % methods heading (mandatory)
   {We made use of X-ray observations with the X-ray telescope 
on-board the \swift\ observatory to refine the X-ray position 
to 3-5\arcsec\ accuracy, and performed 0.1--10~keV spectral analysis. 
We then searched the online catalogues  (e.g. NED, SIMBAD, 2MASS, 2MASX, and 
NVSS) to search for counterparts at other wavelengths. }
  % results heading (mandatory)
   {For all sources, we give a refined X-ray position, provide X-ray spectral 
parameters, identify infrared counterparts, and give magnitudes at optical and 
ultra violet wavelengths seen with UVOT when observations are available.  
We confirm the nature of six sources formerly suspected to be AGN (IGR~J02343+3229, 
J13149+4422, J14579$-$4308, J16385$-$2057, J18559+1535, J19378$-$0617). 
 \correc{Our analysis first leads us to suggest that IGR~J09523$-$6231 and 
IGR~J10147$-$6354 are AGN. While the former has recently been confirmed as
 a Seyfert 1.5 AGN, we suggest
the latter is a Seyfert 2.} All other sources 
may be Galactic sources, in which case their spectral shape may suggest that
they are X-ray binaries. In one case (IGR J19308+0530),  the Galactic nature 
is confirmed through the identification of an F8 star as the counterpart. 
We favour a distance to the source not greater than 1~kpc. The source 
is likely to be a neutron star XRB or a CV. We also report the 
discovery of six serendipitous sources  of unknown nature.}
  % conclusions heading (optional), leave it empty if necessary 
   {}


   \maketitle
%
%________________________________________________________________

\section{Introduction}

Since its launch on October 17, 2002, the INTErnational Gamma-Ray
Astrophysical Laboratory ({\it{INTEGRAL}}, \citet{winkler03}) has
detected about 250 sources that had previously never been seen or 
serendipitously detected once, and not studied. This has been made
possible mainly thanks to the high sensitivity, the wide field of view 
(FOV) and high imaging resolution of the IBIS Soft Gamma-Ray Imager 
(ISGRI, \citet{lebrun03}) sensitive in the 20--300~keV energy range.
These sources 
are named with the acronym IGR J{\tt{RA}}$\pm${\tt{Dec}} and 
we will hereafter refer to them as IGRs\footnote{An up-to-date online
catalogue of all IGRs can be found 
at http://isdc.unige.ch/$\sim$rodrigue/html/igrsources.html}. 
Understanding the nature 
of those sources has a great importance and implication for 
several astrophysical questions. The identification of sources allows
us to perform statistics on families of sources and source population 
studies, while individual studies of new sources allows us to 
better understand the physics of the emission of high energy radiation.
This, in turn, can help us answer questions regarding the evolution of 
stars, galaxies, and/or have cosmological implications while studying Galactic
or extra-galactic sources such as AGN and other quasars.\\
\indent In a recent paper, \citet{bodaghee07} collected all the known 
parameters (such as the absorption column density \nh, or the pulse period 
for Galactic sources, the redshift for AGN, etc.) of all sources detected 
by \integral\ during the first four years of activity,  
and tried to understand the different families of sources by testing
and  searching correlations between those parameters. Their catalogue, 
however, contains  a large number of sources whose high 
energy position is accurate just at the arcmin level, the best 
accuracy achievable with \integral/ISGRI. In most cases this level of accuracy 
is not sufficient to unveil the nature of the source through the 
identification of counterpart at other wavelengths. In some cases,  
a tentative identification is given, mainly when an AGN is found 
within the \integral/ISGRI error box, but this is far from being secure 
as other possible counterpart usually lie in the few arcmin ISGRI error
box.\\
\indent Since the discovery of the first source by \integral, secure 
identification has been possible only through follow-up observations 
with softer X-ray telescopes, either by refinement of the X-ray position 
and identification of the optical/infrared counterpart 
\citep[e.g.][]{rodriguez2006,tomsick06,chaty07}, or by discovery 
of X-ray pulsations in the case of pulsars \citep[e.g.][]{lutovinov05}.
Here, we report the results of several pointed observations made with the \swift\ 
observatory \citep{gehrels04}, on a total of 12 sources that lacked precise 
localisation at soft X-ray energies. We start by introducing the observations
and method of data reduction, before giving the results on each source, i.e., 
refined position, identification of the counterpart, and X-ray spectrum. 
We discuss and summarise the results in the last part of the paper.

\section{Observations and data reduction}
Among all the \swift\ pointed observations of IGRs,  we first restricted 
 our analysis to  sources whose fine (less than$\sim 10$\arcsec) 
soft X-ray position  was not published anywhere. We, then, dis-regarded the 
sources that were not detected in single pointings, because a non detection can have 
several reasons (absorption, under-exposures, variability, transience, etc.) 
that does not help in unraveling the type of a source. We also rejected 
the sources that were not observed in photon mode by the XRT, as no fine position 
can be obtained. In the remaining sample, we focused on sources for which 
either a secure identification was not given in \citet{bodaghee07} or 
those for which we found a mismatch between the X-ray position and the proposed 
association. Our analysis contains a sample of twelve sources. The observing log for 
these sources is reported in Table~\ref{tab:log}. 
All our results are first based on observations made with the XRT 
\citep{burrows05} and UVOT \citep{roming05} telescopes onboard the \swift\ 
observatory. The XRT is a focusing X-ray telescope with an effective area of 110
cm$^2$, and a FOV of about 23\arcmin. It has an imaging resolution 
of 18\arcsec\  between 0.1 and 10 keV, and has a location accuracy for 
point source as low as $\sim$3\arcsec . 

\begin{table}[htbp]
\caption{\label{tab:log} Log of the \swift\ observations 
analysed in this paper. }
\begin{tabular}{lllll}
\hline\hline
Name & \swift\ Sequence & Observation & Exp.$^\star$  & offset$^\dagger$ \\
(IGR) & Number          &  Date     &  (s)     &  (\arcmin) \\
\hline
\hline
J02343+3229 &  00037105001 & 2007-07-03 & 2481 & 5.4 \\
     &  00037105002 & 2007-07-04 & 4265 & 5.4 \\
     &  00037105004 & 2007-07-07 & 6402 & 5.1 \\
     &  00037105006 & 2007-07-07 & 6821 & 5.1 \\
J09523-6231 &  00030927001 & 2007-05-12 & 1027 & 1.9 \\
            &  00030927002 & 2007-06-08 & 2123 & 1.0 \\
            &  00030927003 & 2007-06-14 & 1867 & 2.4 \\
J10147-6354 &  00037048001 & 2007-11-01 & 4538 & 0.8 \\
J11187-5438 &  00037051001 & 2007-09-26 & 6905 & 7.0 \\
            &  00037051002 & 2007-09-30 & 16062 & 2.9 \\
J13149+4422 &  00037093001 & 2007-06-03 & 11493 & 7.8 \\
            &  00037093002 & 2007-09-19 & 1228 & 4.8  \\
            &  00037093003 & 2007-09-20 & 2449 & 7.1  \\
J14579-4308 &  00036621001 & 2007-09-21 & 10725 & 8.0 \\
            &  00036621002 & 2007-09-25 & 6401  & 7.8 \\
            &  00036621003 & 2007-09-25 & 4552  & 7.4 \\
J16385-2057 &  00036649001 & 2007-10-07 & 4554 & 6.1  \\
            &  00036649002 & 2007-10-08 & 4606 & 6.3  \\
J18490-0000 &  00035092001 & 2006-03-05 & 8432 & 3.0  \\
            &  00035092002 & 2006-03-09 & 3828 & 2.0  \\
J18559+1535 &  00036651001 & 2007-07-20 & 3848 & 2.5  \\
            & 00036651002 & 2007-07-29 & 1309 & 2.2 \\
            &  00036651003 & 2007-10-03 & 593 & 4.0\\
            &  00036651004 & 2007-10-18 & 5130 & 6.5 \\
J19308+0530 &  00035357002 & 2006-04-13 & 2120 & 1.3 \\
            &  00035357004 & 2006-05-19 & 4477 & 1.2 \\
J19378-0617 &  00036652001 & 2007-08-07 & 408 & 7.2\\
            &  00036652003 & 2007-09-26 & 5713 & 4.3 \\    
J23524+5842 &  00037065001 & 2007-06-09 & 5901  & 0.9 \\
\hline
\end{tabular}
\begin{list}{}{}
\item[$^{\star}$]XRT exposure time 
\item[$^\dagger$]Offset between 
the direction of the \swift\ pointing and the best \integral\ position
for the given source.
\end{list}
 \end{table}

The XRT data were reduced within the {\tt{HEASOFT}} package v. 6.3.2.
We produced level 2 data with the {\tt{xrtpipeline}} v0.11.5 that 
processes the raw data to obtain clean data products, i.e. 
images, spectra and light curves used for the scientific analysis.
For each pointing, we estimated the X-ray position of the X-ray sources 
with the task {\tt{xrtcentroid}} v.0.2.7. In the various cases where several 
pointings were available for the same source the final position is the mean 
of all positions obtained from the individuals pointings. As the 
error computed by this task includes various effects (especially 
some systematic effects and satellite misalignment), the error we 
report here is also the average of all individual errors. \\
\indent Spectra and light curves were extracted with {\tt{xselect}} v2.4. 
The source spectra and light curves were obtained from a circle of 20 pixels 
($\sim47$\arcsec) radius centred on the best source position. This 
ensures that 90\% of the PSF is enclosed in this region. Background 
spectra and light curves were extracted from a region of the detector 
free of sources and of 40 pixels radius. Exposure maps were produced 
with {\tt{xrtexpomap}} and summed within {\tt{XIMAGE}}. The ancillary response files 
were generated with the tool {\tt{xrtmkarf}} v0.5.5, and corrected with
 the exposure maps at the position of the source. The resultant spectra 
were rebined so as to at least have 20 counts per channel allowing 
the chi-statistics to be used in {\tt{xspec}} v11.3.2ag. If  
this criterion was not achievable, the Cash-statistic was used instead.
In the cases where several pointings are available the spectra were 
averaged together unless large variability was seen in the light curve. The 
spectra were fitted between 0.5~keV and $\sim8$~keV depending on the quality of the 
high energy bins.\\
\indent The UVOT is an UV/optical telescope, whose design is based 
on the OM onboard the {\it{XMM-Newton}} observatory. It has a 
17\arcmin$\times$17\arcmin\ FOV with an angular resolution of about 2\arcsec 
depending of the filter used. It can observe a given field through several
filters, or grisms to perform spectroscopy. The UVOT data (when available) discussed 
in this paper were obtained through one 
or more of the following filters: V (5000--6000~\AA), B (3800--5000~\AA),  
U (3000--4000~\AA), UVW1 (2200--4000~\AA), UVM2 (2000--2800~\AA) and UVW2 
(1800--2600~\AA).  The UVOT individual exposures of a single observation
were summed with {\tt{uvotimsum}} while magnitudes and upper limits were estimated 
at the best X-ray position obtained with XRT with {\tt{uvotsource}} as 
explained in the UVOT analysis 
thread\footnote{http://swiftsc.gsfc.nasa.gov/docs/swift/analysis/threads/uvot\_threads.html}
 by comparison with a region free of sources taken as background reference. Note that, 
as indicated in the UVOT analysis threads, regions of, respectively, 6\arcsec\
radius for the U, B, and V filters and 12\arcsec\ radius, for the UVW1, UVW2, UVM2 filters,
 where given as an input to {\tt{uvotsource}} for the computation of the magnitudes.
The typical errors  are 0.2 magnitudes.

\section{Results}
The main results, X-ray position, presence 
of an infrared counterpart and its type are reported in 
Table~\ref{tab:res}, the IR, optical magnitudes found in the literature, and 
the magnitudes obtained from the analysis of the UVOT data when available are 
reported in Table~\ref{tab:mag}. The spectral results are reported in 
Table~\ref{tab:specres}. \correc{All errors are given at the 90\% confidence 
level.}  The luminosities reported in Table~\ref{tab:specres} are estimated 
at a distance of 1~kpc for sources of unknown type. For the confirmed AGNs, 
we used the redshift reported in Table~\ref{tab:res} and H$_0=65$ km~s$^{-1}$~Mpc$^{-1}$ 
to estimate it. Note that for all sources, the \swift/XRT position 
falls well within the 90\% confidence \integral\ error box. In all cases
there is a single source within the \integral\ error box, and except 
where further discussed below, there are no other bright sources in the FOV, 
although there may be some slight excesses in some of the fields.
The infrared (IR) counterparts are either found in 
the 2MASS point source catalogue, or the 2MASX extended source catalogue. 
Below, we discuss some individual properties for each source.

\begin{table*}[htbp]
\caption{X-ray position (equatorial and Galactic), presence of an IR counterpart and 
reddshift of the 12 sources studied with \swift/XRT.}\label{tab:res}
\begin{tabular}{lllcllccl} 
\hline\hline             
Name & RA & DEC & Error & l & b & Counterpart & Type & Comment\\
(IGR)  & (J2000) & (J2000) & (\arcsec) & (\degr) & (\degr) & (infrared)  &  & \\ 
\hline
\hline
J02343+3229 & 02h 34m 19.9s & +32\degr 30\arcmin 20\arcsec & 3.8 & 146.865 & $-25.540$ & yes/extended & Sey 2 & z=0.016\\
J09523$-$6231 & 09h 52m 20.5s & $-62$\degr 32\arcmin 37\arcsec &4.4 & 283.854 & $-6.507$ &  yes  (Fig.~\ref{fig:09523Charts}) & Sey 1.5  & z=0.252$^\star$\\
J10147$-$6354 & 10h 14m 15.2s &  $-63$\degr 51\arcmin 50\arcsec&4.2 & $286.638$ & $-6.083$ &  yes/point  & Sey 2 (?) &\\
J11187$-$5438 & 11h 18m 21.1s & $-54$\degr 37\arcmin  32\arcsec& 3.7& 289.640 & +5.826 &  yes/point & ? & \\
J13149+4422 & 13h 15m 17.3s &+44\degr 24\arcmin 26\arcsec&3.8 & 108.983 & +72.069 &  yes/extended &  Sey 2 & z=0.035 \\
J14579$-$4308 & 14h 57m 41.3& $-43$\degr 07\arcmin 57\arcsec& 3.8 & 326.114 & +13.984 &  yes/extended & Sey 2 & z=0.016 \\
J16385$-$2057 & 16h 38m 31.1s &  $-20$\degr 55\arcmin 25\arcsec &3.6& 357.728 & +17.020& yes/extended & Sey 1 & z=0.027\\
J18490$-$0000 & 18h 49m 01.6s & $-0$\degr 01\arcmin 18\arcsec & 3.7 & 32.646 & +0.510 & faint IR/point & XRB (?) & \\
J18559+1535 & 18h 56m 00.6s & +15\degr 37\arcmin 58\arcsec & 3.6 & 47.409 & +6.072 &  yes/point & Sey 1 & z=0.084\\
J19308+0530 &  19h 30m 50.9s&  +05\degr 30\arcmin 57\arcsec &   4.3 & 42.381 & $-6.186$ &  yes/point & neutron star XRB or CV & F8 star\\
J19378$-$0617 & 19h 37m 33.1s &  $-06$\degr 13\arcmin 04\arcsec & 3.5 & 32.591 & $-13.074$&  yes/extended & Sey 1.5 & z=0.011\\
J23524+5842 & 23h 52m 22.0s &  +58\degr 45\arcmin 31\arcsec & 4.5 & 115.323 & $-3.238$ & yes/point & ? & \\
\hline 
\end{tabular}
\begin{list}{}{}
\item[$^{\star}$]From \citet{masetti08}
\end{list}
\end{table*}
\begin{table*}
\caption{IR, Optical and UV apparent magnitudes obtained from the literature and online
catalogues, and, when available, from the analysis of the  UVOT data. Upper 
limits are given at the 5-$\sigma$ level.}
\centering
\begin{tabular}{llll}
\hline
\hline
Source & \multicolumn{3}{c}{magnitudes} \\
(IGR)  & IR & Optical & UV               \\
\hline
\hline
J02343+3229 & J=10.021, H=9.184, K=8.772 & B=13.7& U=$17.7$, UVW2=$18.3$\\
J09523$-$6231 &  & & U=$18.6$, UVM2$>19.3$, UVW1$>19.0$\\
J10147$-$6354 & J=14.831, H=13.874, K=12.53 & & UVW2=$16.3$\\
J11187$-$5438 & J=15.534, H=14.869, K=14.455 & &           \\
J13149+4422 & J=12.248, H=11.624, K=10.821 & B=16.5 & UVM2=$16.9$,UVW1=$16.7$\\
J14579$-$4308 & J=10.745, H=9.918, K=9.578 & B=15  & \\
J16385$-$2057 & J=12.527, H=11.644, K=11.106 & &\\
J18490$-$0000 & K=14.159 &V$>20.2$ & \\
J18559+1535   & J=13.639, H=12.584, K=11.438 & & UVM2$>19.4$, UVW2$>20.2$\\
J19308+0530   & J=9.617, H=9.245, K=9.130 & V=11.0, B=11.3 & U$<11.9$, UVM2=12.5, UVW2=13.7 \\
J19378$-$0617 & J=10.88, H=19.141, K=9.666 & V=15.35, B=16.12 & UVW1=14.4 \\
J23254+5842   & J=16.127, H=15.059, K=13.9155 & & U$>20.7$\\
\hline
\end{tabular}
\label{tab:mag}
\end{table*}
\subsection{\object{IGR~J02343+3229}}
IGR~J02343+3229 was discovered with \integral\ by  
\citet{burenin06} and \citet{Krivonos07}, and promptly associated with NGC 973 a
likely Sey 2 AGN \citep{burenin06}.  
The average XRT position is 3.7\arcsec\ away from the reported position 
for NGC 973. The angular size of the galaxy is 
3.98\arcmin, which renders the association of the 
X-ray source and the Galaxy very likely. 
We found a possible counterpart in the 2MASX catalogue of extended  
sources. 2MASX J02342010+3230200 is 2.0\arcsec\ away from the centre of 
the \swift/XRT error box. Both objects fall 
well within the XRT error, further increasing the probability 
that IGR J02343+3229 is an AGN. This AGN is at  z=0.016 as reported in 
NED from various sources. 
We examined the UVOT data of the third exposure that contains
coverage in the U and UVW2 filters. The best XRT position of IGR~J02343+3229
clearly contains the nucleus of a galaxy detected in both filters. The magnitudes
obtained at the best XRT position within the recommended 6\arcsec and 12\arcsec radius 
regions of extraction are reported in Table \ref{tab:mag}.\\
\indent We remark the presence of an additional source and another possible faint excess 
in the field. Both features are outside the   
\integral\ error box. The position of the source 
(named \object{SWIFT J023405.1+322707}) is reported in Table~\ref{tab:pos}. 
It lies 2.5\arcsec\ from HD 15896 (=2MASS J02340529+3227074)
 a K0 star, with B=8.74, V=7.65, J=5.754, H=5.259, K=5.087, which 
may suggest the source has a Galactic origin.  \\
\indent We combined the four observations to perform a spectral analysis 
of IGR~J02343+3229.
The average spectrum has 1437 cts for a total of 19910~s exposure. An absorbed 
power law fits the data well with a reduced $\chi^2$ (hereafter $\chi_\nu^2$) of 
1.11 for 63 degrees of freedom (DOF). The parameter values are reported in 
Table~\ref{tab:res}. The value of \nh\ is about 40 times higher than the mean 
absorption along the line of sight, which indicates that IGR~J02343+3229 is 
an intrinsically absorbed AGN. Note that intrinsic absorption is expected in 
the case of a Sey 2 AGN.

\subsection{\object{IGR~J09523$-$6231}}
\begin{figure}
\epsfig{file=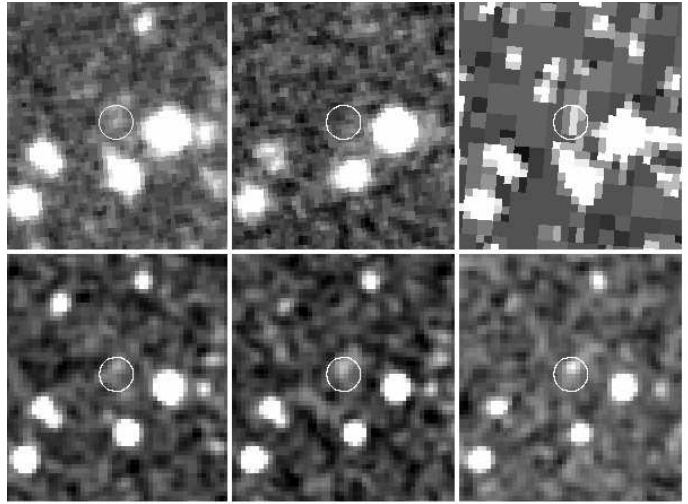,width=\columnwidth}
\caption{$\sim46$\arcsec$\times59$\arcsec\ finding charts of the field 
around IGR J09523$-$6231. From left 
to right and top to bottom DSS II infrared, red, blue images, and 2MASS 
J, H and K-band images. In all images, the small circle represents the 
\swift/XRT error box. }
\label{fig:09523Charts}
\end{figure}

IGR~J09523$-$6231 was first reported in the third edition of the IBIS
catalogue \citep{bird07}. The source was only detected in the 20--40~keV 
energy range at a low significance of 5.3$\sigma$ over 290 ks of 
observation \citep{bird07}.
Only one X-ray source is found in the \integral\ error box. It is, however, 
 extremely weak  and the positions obtained from the three \swift/XRT 
pointings can differ by few arcsec (up to 4). 
There is nothing reported in the various online catalogue at less 
than 10\arcsec\ from the average X-ray position.
Fig. \ref{fig:09523Charts} shows the DSS II R-band and infrared, and the 
2MASS J, H, and K-band  images. 
As can be seen in Fig.~\ref{fig:09523Charts}, a weak source is found within 
the \swift\ error box in the 2MASS and DSS II images. 
The source may even be blended or extended as seen from the infrared images. 
We examined the UVOT data for the presence of a possible counterpart at the 
position of the source. A faint source is visible in the U-band only 
(Table~\ref{tab:mag}).\\
\indent We remark the 
presence of another  X-ray source in the \swift/XRT FOV. It is well outside the 
\integral\ error box. Its position is reported in Table~\ref{tab:pos}. This additional 
source (named \object{SWIFT J095238.4-622316}) is within the 
14\arcsec\ {\it{ROSAT}} error box of 1RXS~J095237.2-622310, which suggests 
the two sources are the same. A counterpart 
with m$_{\mathrm{UVM2}}=17.2\pm0.2$ is found in the XRT error 
box of SWIFT J095238.4-622316. This source is unfortunately outside the 
UVOT FOV of the pointings made with the U and UVW1 filters. \\
\indent We combined the three observations to perform a spectral analysis.
The  spectrum has 215 cts for a total of 5017~s exposure. 
An absorbed power law fits the data well ($\chi_\nu^2$=0.9 for 6 DOF). 
The fitted value of \nh\ is about 40 times higher than the average value of 
the absorption along the line of sight, which indicates that IGR~J09523$-$6231
is intrinsically absorbed. The detection of the source in the U-filter
and the high X-ray absorption \correc{first led us to think the object 
was a Sey 2 AGN, since in those objects, 
the optical emission is thought to be produced outside of the absorbing 
matter affecting the X-ray emission. This despite the position of the source 
close to the Galactic plane (Table \ref{tab:res}) which may have rather pointed towards 
a Galactic source, as AGN tend to statistically be found at high Galactic latitudes. 
In a very recent work, \citet{masetti08}, showed
through optical spectroscopy, that IGR~J09523$-$6231 was indeed an AGN. They 
refined the type to a Sey 1.5.}

\subsection{\object{IGR J10147$-$6354}}
IGR~J10147$-$6354 was first reported by 
\citet{bird07}. The source was detected in the 40--100~keV 
energy range at a significance of 4.9$\sigma$ during 1340~ks of observation.
The best XRT position is well within the ISGRI error box of 5\arcmin.\\
\indent There is a single source in the 2MASS point source catalogue within
5\arcsec\ from the XRT position. 
2MASS J10141554-6351500 is at 2.1\arcsec\ away from the centre of the \swift\ 
error box. It is also well 
detected in the UVOT UVW2 filter. \\
\indent The XRT spectrum has 93 cts for a total exposure of 4538~s. 
An absorbed power law fits the data well, with a C-statistic 
value=32.6 for 29 bins. 
The value of \nh\ is a factor about 7.3 times higher than the average 
value of the absorption along the line of sight which indicates that 
the absorption is mostly intrinsic to the source. The detection of a bright
UV counterpart, as for the former source,  suggests either a very close Galactic
or a far and bright extragalactic object. Again, the low Galactic latitude 
would rather tend to point towards a Galactic source rather than 
an extragalactic one.  Dereddening the magnitudes with the 
absorption on the line of sight does, however, not lead to any known stellar 
spectral type for a Galactic source. This, the detection in the UV-band,  
and the intrinsic absorption of the source would tend to favour a Sey 2 AGN.
\begin{table}
\caption{Spectral parameters of the 12 sources studied with \swift/XRT. 
}\label{tab:specres}
\begin{tabular}{lllll} 
\hline\hline             
Name & \nh & $\Gamma$ & Flux$^\star$ & Lumin.$^\dagger$ \\
(IGR)  & $\times10^{22}$ \cm2 & &  & \\ 
\hline
\hline
J02343+3229   & $2.2\pm0.4$ & $1.3\pm0.2$ & 1.0$\times10^{-11}$ & 6.7$\times10^{42}$\\
J09523$-$6231 & $8_{-4}^{+5}$ & $2.3_{-1.1}^{+1.3}$ & 9.1$\times10^{-12}$ & 1.5$\times10^{43}$\\
J10147$-$6354 & $2.0_{-1.1}^{+1.6}$ & $1.7_{-0.8}^{+0.9}$ & 2.1$\times10^{-12}$ & 2.5$\times10^{32}$\\
J11187$-$5438 & $0.28_{-0.07}^{+0.08}$ & $1.5\pm0.14$ & 3.4$\times10^{-12}$  & 4.1$\times10^{32}$\\
J13149+4422   & $5.2_{-1.0}^{+1.5}$ & $1.7_{-0.3}^{+0.4}$ & 1.2$\times10^{-11}$ &  4.1$\times10^{43}$\\
J14579$-$4308 & $20\pm4$& $2.9\pm0.6$ & 1.8$\times10^{-11}$  & 1.2$\times10^{43}$\\
J16385$-$2057 & 0.21$\pm0.04$ & $2.1\pm0.1$ & 7.7$\times10^{-12}$ & 1.4$\times10^{43}$\\
J18490$-$0000 & 5$\pm2$ & 1.8$_{-0.6}^{+0.7}$ & 6.4$\times10^{-12}$ & 7.6$\times10^{32}$\\
J18559+1535 &  $0.7\pm0.1$ & 1.6$\pm0.1$ & 1.4$\times10^{-11}$ & 2.5$\times10^{44}$\\
J19308+0530 &   $<0.3^\ddag$ & 3.0$_{-0.5}^{+1.4}$ & 8.2$\times10^{-14}$ & 9.8$\times10^{30}$\\
J19378$-$0617 &  0.15$\pm0.05$ & 2.5$\pm0.2$ & 4.4$\times10^{-11}$ & 1.3$\times10^{43}$\\
J23524+5842 &  $6_{-2}^{+4}$ & 2 {\it{frozen}} & 2.9$\times10^{-12}$  & 3.4$\times10^{32}$\\
\hline 
\end{tabular}
\begin{list}{}{}
\item[$^{\star}$]2--10~keV unabsorbed (\ergcms)
\item[$^\dagger$]2--10~keV unabsorbed luminosity (erg~s$^{-1}$)
\item[$^\ddag$]90$\%$ upper limit
\end{list}
\end{table}
\subsection{\object{IGR J11187$-$5438}}
IGR~J11187$-$5438 was first reported by \citet{bird07}. The source was detected 
at a 18--60~keV significance of 6.3$\sigma$ during 1016~ks of observations.
There is a single source in the 2MASS point source catalogue within
5\arcsec\ from the XRT position. 2MASS J11182121-5437286 is at 
3.5\arcsec\ away from the centre of the \swift\ error box. The source is also 
visible in the DSS II IR and R-Band images, 
although it is quite weak in the latter. The UVOT was not 
operating during either of the observations.\\
\indent We combined the two observations to perform a spectral analysis.
The average spectrum has 1328 cts for a total of 21888~s exposure. 
An absorbed power law fits the data well ($\chi_\nu^2=0.91$ for 55 DOF). 
The value of the absorption is compatible with 
the  Galactic absorption along the line of sight, which may mean that 
the source is not intrinsically absorbed. 

\subsection{\object{IGR J13149+4422}}
IGR J13149+4422 was first reported in \citet{sazonov07}  and 
\citet{Krivonos07}. The latter give an IBIS position of 
\ra=13h 14m 58s \dec=+44\degr 23\arcmin\ with a 1-$\sigma$ uncertainty as 
high as 2.1\arcmin.  It was then tentatively identified with 
Mrk 248 a Sey 2 AGN at z=0.037 \citep{sazonov07}. 
The best \swift/XRT position is  3.56\arcmin\ from the IBIS position, therefore 
within the 90$\%$ confidence error box. It is compatible with the best reported position 
of UGC 8327 NED02 (at \ra=13 h15m 17.270s, \dec=+44\degr 24\arcmin 25\farcs60 according 
to the latest measurements available in  NED, thus at 0.13\arcsec\ from the centre of 
the \swift\ error box). Note that this is also at 16\farcs8 from the position reported in 
\citet{bodaghee07}, and in SIMBAD for Mrk 248. We remark that SIMBAD reports 
an earlier measurement than NED, and also mentions Mrk 248 as being UGC 8327, a pair 
of interacting galaxies (UGC 8327). NED, on the other hand, associates Mrk 248 
with one of the components of the pair of interacting galaxies (UGC 8327 NED02).
Mrk 248 is also reported in the  2MASS extended source catalogue as 
2MASX J13151725+4424259 whose best position is at only 0.64\arcsec\ away from the 
best X-ray position. The 2MASS source is also classified as an emission line galaxy in 
SIMBAD. An extended or blended source is also
clearly visible in the UVOT UVM2 and UVW1 images within the XRT error box.
The spatial coincidence between all sources likely indicates that IGR~J13149+4422 
is a Sey 2 AGN.  \\
\indent We extracted a  spectrum from three pointings. The spectrum
has 964 cts for a total of 15116~s. It is well fitted by an absorbed power law
($\chi^2$=1.11 for 32 DOF). We note the presence of positive residuals at low energy.  
The value of \nh\ is about 
430 times higher than the average absorption along the line of sight, which 
indicates that the source is an intrinsically absorbed AGN as would be 
expected in the case of a Sey 2 AGN.

\subsection{\object{IGR J14579$-$4308}}
\begin{figure}[htpb]
\epsfig{file=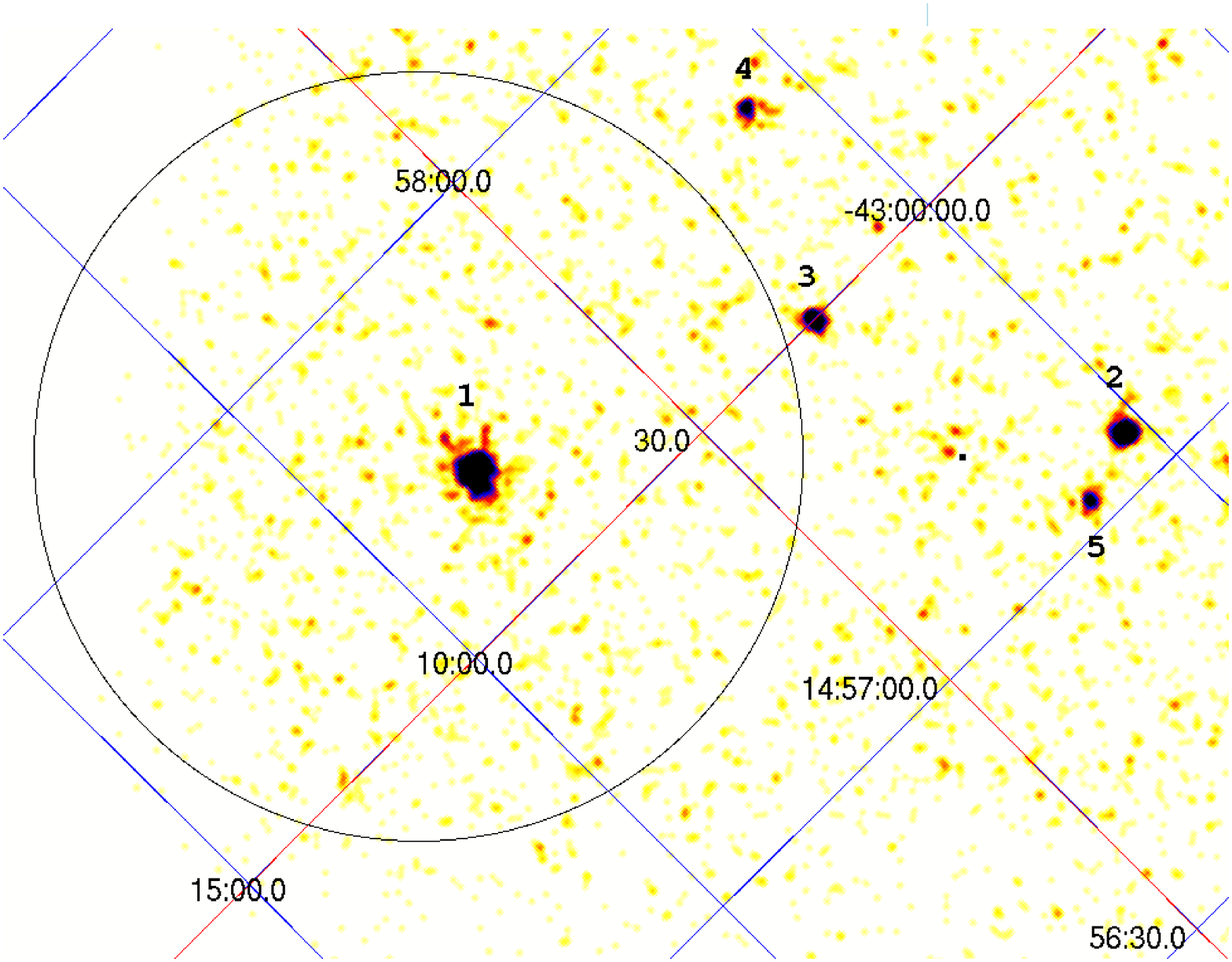,width=\columnwidth}
\caption{\swift\ XRT image of the field around 
IGR J14579$-$4308. The 5 sources are indicated with numbers. Source 1 is 
the X-ray counterpart to IGR J14579$-$4308. The big circle is the maximum 
IBIS error box of 6\arcmin\ reported in \citet{revnivtsev05}.}
\label{fig:14579}
\end{figure}

IGR J14579$-$4308 was first reported by \citet{kalemci05} from an 
IBIS/ISGRI observation of SN 1006.  It was promptly suggested to be 
an AGN due to its positional coincidence with 
VV 780 a Sey 2 AGN. \citet{revnivtsev05} later reported the presence of the source 
from a survey of the Galactic Crux arm tangent. 
The XRT images show that the field is crowded with about five bright 
sources in the field of view (Fig. \ref{fig:14579}). Only the brightest 
source is, however,  
found within the maximum of 6\arcmin\ error box of ISGRI  \citep{revnivtsev05}. 
It is labeled 1 in Fig. \ref{fig:14579}.
The best XRT position is 42\arcsec\ away from the reported position
of IC 4518/ VV 780 which rules out an association between the X-ray source 
and the AGN. \citet{bird07} suggested a possible association of the 
INTEGRAL source with IC 4518A, a Sey 2 galaxy as inferred from optical spectroscopy
\citep{masetti07c}, also the western component of a pair of 
interacting galaxies named MCG-07-31-001. The best \swift\ position is, indeed, 
just 1.48\arcsec\  away from the position of IC 4518A (also named VV 780 NED01) 
as reported in NED.  Our \swift\ analysis, therefore, 
strengthens the association of both objects, and we confirm that IGR~J14579$-$4308 
is a Sey 2 galaxy.  Note that, as for IGR J13149+4422, the position of 
IGR J14579$-$4308 reported in \citet{bodaghee07} and in SIMBAD is that of the pair of
interacting galaxies, while we identify here one of the components of this double system 
as the best counterpart to the \integral\ source thanks to the refined position. 
The best positions  for the other sources are reported in Table~\ref{tab:pos}. Only
source 2 on Fig.~\ref{fig:14579} has an infared counterpart reported in 2MASS.
2MASS J14570433$-$4300187 indeed lies 1\farcs5 from \object{SWIFT J145704.4$-$430020}. 
No UVOT data are available from any of the observing sequences.

 We combined the three observations to perform a spectral analysis.
The average spectrum of IGR J14579$-$4308 has 703 cts for a total of 21685~s exposure. 
A simple absorbed power law does not fit the data well ($\chi_\nu^2=2.88$ for 29 DOF).
A large excess is visible below 2 keV. Adding a black body to the model improves
the fit to $\chi_\nu^2=1.25$ for 27 DOF. The black body radiation is  not 
absorbed and has a temperature of $0.28_{-0.06}^{+0.11}$~keV. 
The best parameters of the other spectral components (power law
and \nh) are reported in Table~\ref{tab:res}. Replacing
the black body by a disc model ({\tt{diskpn}} in XSPEC) provides a good description of the 
spectrum ($\chi_\nu^2=1.28$ for 27 DOF). The inner disc temperature is 
0.5$_{-0.2}^{+0.7}$~keV. The high value of the absorption 
indicates that the object is intrinsically highly absorbed,  
as would be expected in a Sey 2 AGN. Further spectral results 
will be reported in Kalemci et al. (in prep.).

\begin{table}
\caption{Best XRT positions of serendipitous sources 
found in this study. \label{tab:pos}}
\begin{tabular}{llll}
\hline
Name & \ra & \dec & error \\
(SWIFT) & & & (\arcsec)  \\
\hline
\hline
J023405.1+322707 &  02h 34m 05.1s & +32\degr 27\arcmin 07\arcsec & 4.1\arcsec\\
J095238.4$-$622316 &  09h 52m 38.4s & $-62$\degr 23\arcmin 16\arcsec & 4.1\arcsec\\
J145704.4$-$430020 &  14h 57m 04.4s & $-$43\degr 00\arcmin 20\arcsec & 3.8\arcsec \\
\object{J145729.9$-$430231} &  14h 57m 29.9s & $-$43\degr 02\arcmin 31\arcsec & 4.0\arcsec\\
\object{J145746.9$-$430056} &  14h 57m 46.9s & $-$43\degr 00\arcmin 56\arcsec & 4.2\arcsec\\
\object{J145702.3$-$430128} &  14h 57m 02.3s & $-$43\degr 01\arcmin 28\arcsec & 4.0\arcsec\\
\hline
\hline
\end{tabular}
\end{table}

\subsection{\object{IGR J16385$-$2057}}
IGR~J16385$-$2057 was first reported by \citet{bird07}. 
Based on a positional coincidence with 
1RXS J163830.9$-$205520 and  Oph J163830$-$2055,  it was suggested to be the 
X-ray counterpart to those objects. Optical spectra allowed \citet{masetti06}
to tentatively classify it as a Sey 1 at z=0.027.
The \swift/XRT position is 2.5\arcsec\ away from the position of 
2MASX J16383091$-$2055246 an AGN at z=0.026 \citep{hwm00}. 
This object is also compatible Oph~J163830$-$2055, and therefore 
the tentative identification as a Sey 1 seems confirmed by the refined 
XRT position. Note that the \swift\ position is also 5.73\arcsec\  
from 1RXS~J163830.9$-$205520, well within the ROSAT error box (7\arcsec).
No UVOT data are available for this source.\\
\indent  We combined the two observations to perform a spectral analysis.
The average spectrum of IGR~J16385$-$2057 has 2195 cts for a total of 9168~s 
exposure. An absorbed power law fits the data well ($\chi_\nu^2=1.27$ for 85 DOF).
The value of the absorption is a factor of two higher than the total 
absorption along the line of sight. This may indicate
that a part of the absorbing material is intrinsic to the object.

\subsection{\object{IGR~J18490$-$0000}}
IGR~J18490$-$0000 was first reported by \citet{molkov03} from a survey of 
the Sagittarius Arm tangent region. Nothing more is known about this source.
There is a single 2MASS point source within the \swift/XRT error box.
2MASS J18490182-0001190	lies 3.55\arcsec\ from the centre of the \swift\ error 
box. It has a well-measured magnitude only in the K-band, while it is not
detected in the UVOT V-filter.\\
\indent 
We extracted an average spectrum from the two pointings. The spectrum has 
441 counts for a total of 12208~s and is  well fitted 
by an absorbed  power law ($\chi_\nu^2=0.40$ for 17 DOF). 
The value of \nh\ is about 3 times higher than the average value of the 
absorption along the line of sight. This may indicate that IGR~J18490$-$0000 
is intrinsically absorbed. This source is likely a Galactic X-ray 
binary because of the presence of a point source K-band counterpart, its spectrum 
intrinsically absorbed and typical of an XRB, and its position towards the
Sagittarius Arm. 
\subsection{\object{IGR J18559+1535}}
The first mention of an \integral\ detection of this source is reported in 
\citet{bird06}, as 2E 1853.7+1534, a Sey 1 AGN. The \swift/XRT position is 
about 12\arcsec\  away from the best reported position for 
2E 1853.7+1534/2MASX J18560128+1538059. 
There are no extended sources or NED objects within 10\arcsec\ of the best position. 
In the XRT  error box there are, however, a 2MASS point source,  
2MASS J18560056+1537584 at 0.7\arcsec\ and a {\it{ROSAT}} source 
1RXH J185600.4+153757 at 2.89\arcsec. The positional coincidence of these 
two sources probably indicates that they are related.  We do not detect 
any source in the UVM2 and UVW2 UVOT filters at the best XRT positions.
\citet{masetti06} and \citet{bikmaev06} performed optical follow-up 
observations of the field of this object. \citet{masetti06} suggested an association of 
this gamma-ray source with the {\it{ROSAT}} source, which allowed them to perform 
optical spectroscopy of the object. They identified it as a Sey 1 galaxy at  
z=0.084 $\pm $ 0.001. \citet{bikmaev06} obtained an optical position 
exactly coincident with the \swift\  position of IGR J18559+1535. They also 
refine the redshift to z=0.0844$\pm$0.0002. Our X-ray refined position strongly confirms
the identification of the high energy source as a Sey 1 galaxy. \\
\indent We extracted an average spectrum from the four pointings. The spectrum has 
1572 cts for a total exposure of 8378~s. The spectrum is well fitted with an 
absorbed power law ($\chi_\nu^2=0.90$ for 68 DOF).  
The value of \nh\ is about twice the value of the absorption 
along the line of sight which could indicate that the object has 
some intrinsic absorption, although at a low level, as would be expected from a Sey 1
 AGN.
\subsection{\object{IGR~J19308$+$0530}}
IGR~J19308$-$0530 was first reported by \citet{bird06}. It was detected at a 
20--60~keV significance of 6.6$\sigma$ for a total of 949~ks of observing time. 
Within the \swift/XRT error box lie a known star, TYC 486-295-1 also reported 
in the 2MASS catalogue as 2MASS J19305075+0530582.
The average \swift/XRT position is just 2.5\arcsec\ away from the best position 
of TYC 486-295-1 = 2MASS J19305075+0530582 a F8 star. 
This positional coincidence may suggest IGR~J19308+0530 is a Low Mass/Intermediate 
Mass  X-ray Binary. However, all DSII and 2MASS images are saturated at the position of  
TYC 486-295-1/2MASS J19305075+0530582, and we do not exclude that this 
is actually a blend of sources.The source saturates the UVOT images
in the U, B and V-filters, and is clearly detected in the other filters 
(Table~\ref{tab:mag}).\\
\indent We extracted an average  spectrum from the two pointings.
The source is quite weak, with a 82 cts spectrum for a total 
exposure of 6556~s. The spectrum is 
well fitted by an absorbed power law (C-statistic value 20.7 for 22 bins).
The value of \nh\ is compatible with the value of the absorption along the 
line of sight. This indicates that the object is not intrinsically  absorbed. As 
can be seen in Table~\ref{tab:res} the power law photon index is quite soft. 
As such a steep power law may be indicative of a thermal spectrum, 
we replaced it by an absorbed absorbed black body. The
 fit is rather good (C-statistic value 26 for 23 bins). The best parameters 
are \nh$<0.15\times10^{22}$\cm2 (90\% upper limit) and kT=0.26$\pm0.5$~keV.
Note that a bremsstrahlung also fits the spectrum well. \\
\indent We dereddened the data with the value of \nh\ obtained from the spectral 
fit (Table \ref{tab:res}) and the value of Galactic value of the absorption along the 
line of sight ($\sim 0.2\times10^{22}$ \cm2). We then estimated the distance to 
the source assuming the optical/infrared counterpart is an F8 star, corresponding 
to the 2MASS source. Typical 
parameters of an F8 star favour a low absorption (0.2$\times10^{22}$ \cm2) at 
a distance not greater than 1~kpc. At this distance the 
2--10~keV unabsorbed luminosity is $\sim4\times10^{31}$ erg~s$^{-1}$ for the 0.2~keV 
black body. 
These parameters suggest that IGR~J19308$+$0530 
is most probably a neutron star XRB in quiescence, or a CV. If it is much closer than 
1~kpc, and although an F8 companion may be a quite extreme case, the source would 
rather be a CV. 

\subsection{\object{IGR J19378$-$0617}}
A first mention of an \integral\ source at this position is given 
in \citet{molkov03}. \correc{They suggested that the source of the hard X-ray emission 
was SS 442 (1H 1934$-$063). }
The authors, however,  suspected the presence of a new source 
given the 5.6\arcmin offset of their position of the IBIS source and that of SS 442. 
IGR~J19378$-$0617 was  then reported by \citet{bird07} at a position slightly 
different than that of \citet{molkov03}. 
It was detected at a 18--60~keV SNR of 5.7$\sigma$. It was promptly suspected to 
be a Sey 1 AGN. Within the \swift/XRT error box lies an extended 2MASS source. 
2MASX J19373299-0613046 is at 1.0\arcsec\ from the best \swift\ position.
Note that this source is also associated with SS 422/1H 1934$-$063 mentioned
in the catalogue of \citet{molkov03}. 
It is a Seyfert 1.5 galaxy at z=0.011. The 2MASS source is also
reported in several catalogues,  while it is clearly detected in the 
UVW1-filter in the UVOT data. 
It is a known radio source (NVSS J193733-061304) and a 
known X-ray source (1RXS J193732.8-061305). \\
\indent We extracted an average spectrum from the two pointings. 
The spectrum has 4222 cts for a total exposure of 2910~s which may indicate
that pile up is not negligible. To avoid pile up effect we extracted the source
spectrum from an annular region centred on the best source position excluding the 
10 central pixels. The outer radius of the annulus was set to 40 pixels. The 
resultant spectrum has 1201 cts. An absorbed power law fits the spectrum well 
($\chi_\nu^2$=1.1 for 45 DOF). The value of \nh\ is compatible with 
the average value of the absorption along the line of sight, which indicates that
IGR J19378$-$0617 is a Sey 1.5 which is not intrinsically absorbed.

\subsection{\object{IGR J23254+5842}}
IGR J23254+5842 was first reported  in the third edition of the IBIS
catalogue \citep{bird07}. It was detected in both the 20--40~keV and 
40--100~keV energy ranges, and the 18--60 significance of its detection 
was 6.3$\sigma$ out of a total of 1780 ks of observations.
Within the \swift/XRT error box lies a single 2MASS source. 2MASS J23522211+5845307 
is 0.92\arcsec\  away from the centre of the \swift\ error box, and it is also bright in 
the DSS I \& II images. Although a  very faint source may be present in the 
U-filter of the UVOT data, we cannot precisely determine its magnitude. \\
\indent We extracted a spectrum from the single \swift/XRT pointing.
The source is very weak with a 70 cts spectrum  for a total of 5902~s exposure.
A simple power law fits the spectrum well (C-statistic value = 19.5 for
13 bins). Adding an absorbing component improves the statistic to 16.3 
for 14 bins.  Note that when all 
parameters are left free to vary they are poorly constrained. We therefore 
froze the power law photon index to 2.
The prefered value of \nh\ may indicate that the absorption is intrinsic
to the source, but the poor quality of the data prevents us any firm 
conclusion on that matter.

\section{Discussion and conclusions}
We analysed \swift/XRT observations of 12 IGRs that previously lacked X-ray 
position at several arcsec accuracy. This lack of fine positions at X-ray energy
either prevented a confirmation of the supposed type of the object or simply
prevented nature of the object to be found. The refinement of the 
X-ray positions allowed us to identify potential counterparts at infrared, 
optical and UV wavelengths for all of them. We also report the detection of 
six serendipitous sources of unknown nature although in the case of 
SWIFT~J023405.1+322707, 
a K0 star is the likely counterpart and thus suggests the source has a galactic origin.
All IGRs that were formerly suspected to be 
AGN were confirmed through our analysis as indeed being so. This shows that 
although the error box of \integral\ can contain several candidate counterparts, 
when an AGN is found inside it is usually also at the origin of the hard 
X-ray emission. This is especially true for sources that have high galactic 
latitude ($>10$\degr). We confirm that IC 4518A is the counterpart to 
IGR~J14579$-$4308, and therefore that this source is a Sey 2. We also 
truly identify IGR~J19378$-$0617 as a Sey 1.5 galaxy, with known infrared, radio 
and X-ray counterparts. In  IGR~J14579$-$4308 we detected a soft excess 
in the X-ray spectrum. Soft excesses have been 
detected in a large number of X-ray spectra of AGNs \citep[e.g.][]{porquet04}. 
The estimated luminosity of this soft excess  
is 7.9$\times10^{40}$~erg~s$^{-1}$, which is compatible with an origin intrinsic
to the AGN. When fitting with a {\tt{diskpn}} model instead of a 
black body, we obtain a lower limit on the inner radius of the 
disk R$_{in}>6$R$_{G}$.\\
\indent For the other sources, we either found in the optical and infrared
surveys, faint sources within the \swift\ error box. In two cases these counterparts 
may be extended or a blend of sources, which prevents an identification to be given.
In two cases U and UVW2 counterparts were found. 
For IGR J09523$-$6231 we first proposed a tentative Sey 2 identification. \correc{The AGN 
nature of the source has recently been confirmed. It  is, however, a Sey 1.5 \citep{masetti08}.} 
For IGR J10147$-$6354 the identification 
as an AGN, possibly a Sey 2, seems more secure. We, however, stress 
that only through optical spectroscopy of the counterpart shall the 
identification be firmly given. For the others, the fact that these objects 
have point sources as optical/infrared counterparts may suggest that they are Galactic
sources, although this is not a definite proof. 
Only in some specific cases this is, however, strongly supported 
by some additional facts. IGR J19308+0530 has an F8 star as the most likely 
counterpart. It spectrum is indicative of little intrinsic absorption (which 
may also suggest that it is a close object), and is very soft. 
IGR J18490$-$0000 has a K-band counterpart. Its spectrum is intrinsically absorbed 
and resembles that of an XRB. Its position in the direction of the Sagittarius Arm tangent
would strengthen its Galactic nature, as the arms of the Galaxy are sites with 
high-density of sources. Note that  these sources lie at Galactic latitudes $<7$\degr\
which may suggest that they are associated with the Galactic Plane, further 
supporting a Galactic origin. In all those cases (but IGR J19308+0530), 
the power law photon index returned 
by the spectral fit may suggest they are XRBs, although a more
definite identification would require optical spectroscopy of the counterpart,
and study of the temporal variability of the X-ray source.\\

\begin{acknowledgements}
JR thanks A. Maury for very useful suggestions on early versions of 
this paper. The authors thank Emrah Kalemci for communications of 
preliminary results on VV 780 prior to publication. We also warmly thank
 N. Masetti and I. Bikmaev for pointing us more precise results concerning some 
of the sources, and the anonymous referee for a very useful report which helped 
us to improve the paper. 
We acknowledge the use of data collected with the \swift\ observatory. This 
research has made use of the SIMBAD database, operated at CDS, Strasbourg, France.
It also makes use of data products from the Two Micron All Sky Survey, which 
is a joint project of the University of Massachusetts and the Infrared Processing 
and Analysis Center/California Institute of Technology, funded by the National 
Aeronautics and Space Administration and the National Science Foundation.
This research has made use of the NASA/IPAC Extragalactic Database (NED) 
which is operated by the Jet Propulsion Laboratory, California Institute of Technology, 
under contract with the National Aeronautics and Space Administration. We acknowledge the use of NVSS, 
DSS  online catalogues. 
\end{acknowledgements}

\end{document}